\documentclass[
   ,final
  ]
  {aipproc}
\usepackage{epsfig}

\def\gsim{\,\lower.25ex\hbox{$\scriptstyle\sim$}\kern-1.30ex%
\raise 0.55ex\hbox{$\scriptstyle >$}\,}
\def\lsim{\,\lower.25ex\hbox{$\scriptstyle\sim$}\kern-1.30ex%
\raise 0.55ex\hbox{$\scriptstyle <$}\,}

\newcommand{\Ftwo}   {\mbox{$\tilde{F}_2$}}

\newcommand{\Fz}   {\mbox{$\tilde{F}_3$}}

\newcommand{\Fem}  {\mbox{$F_2$}}

\newcommand{\Fint} {\mbox{$F_2^{\gamma Z}$}}
\newcommand{\Fwk}  {\mbox{$F_2^{Z}$}}

\newcommand{\Fzint} {\mbox{$F_3^{\gamma Z}$}}
\newcommand{\Fzwk}  {\mbox{$F_3^{Z}$}}

\newcommand{\QQ}  {\mbox{${Q^2}$}}

\layoutstyle{6x9}

\SetInternalRegister\hbadness{8000} 

\newcommand\doingARLO[2][]{%
  \ifx\mmref\undefined #1\else #2\fi
}

\begin{document}

\title{The Physics of Deep-Inelastic Scattering at HERA}


\author{Cristinel Diaconu}{
  address={Centre de Physique des Particules de Marseille  and \\Deutsches Elektronen Synchrotron, Notkestrasse 85, 22607 Hamburg, Germany},
  email={diaconu@cppm.in2p3.fr},
}


\begin{abstract}
In this paper an introduction to the physics of deep-inelastic scattering is given together with an account of some of the most recent results on the proton structure 
obtained in electron-- and positron--proton collisions at the HERA collider.

\end{abstract}

\date{\today}

\maketitle


\section{Introduction}

The investigation of the matter structure using particle collisions started in early XX century with  Geiger, Mardsen and Rutherford discovery of the atomic nucleus using $\alpha$-particle scattering on a gold foil. The principle of the measurement is related to the spatial resolution obtained using high energy particles.
The `resolving power'
can be expressed as $\delta=\sim 200~{\mathrm MeV}/ Q$~$[$10$^{-15}$~m$]$ and
is related to the uncertainity principle: the higher the transfer momentum (denoted by $Q$), the smaller the details that can be "flashed" and imprinted in the distribution of the scattered (point-like) particle.
The search for further substructure levels continued with the scattering of leptons on light nuclei ($H$,$D$) in order to investigate the structure of protons and neutrons, the main components of the nuclear matter. The finite size of the proton was established in elastic scattering and its components,  the quarks and the gluons, were discovered using  electrons of higher and higher energies, which were ultimately  able to break the nucleon in the so called deep inelastics scattering (DIS).  In this paper, an introduction~\cite{disbooks} to the physics of DIS is given
 using as examples recent measurements of the proton structure performed at HERA electron-proton collider\footnote{The paper is based on an introductory lecture presented at the ``Carpathian Summer School of Physics on Exotic Nuclei and Nuclear/Particle Astrophysics'' , August 20-31, 2007, Sinaia, Romania.
}. 


\section{The observables of lepton--hadron scattering}
\label{sec:disreco}
\begin{figure}[hhh]
\begin{minipage}[t]{0.35\linewidth}
  \begin{center}
   \mbox{\epsfig{file=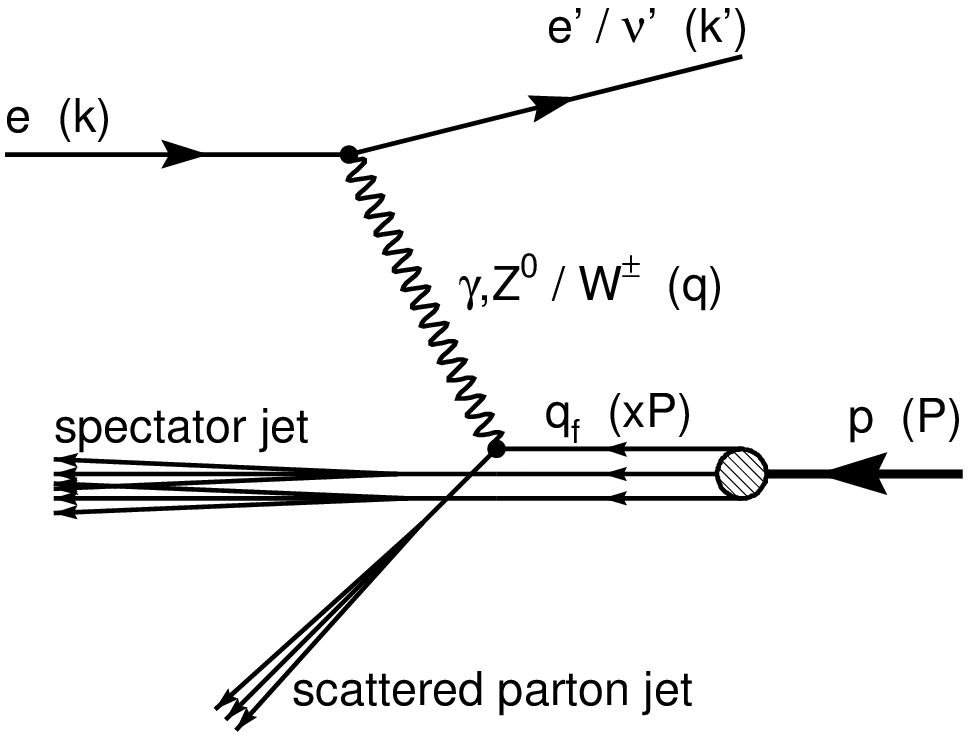,width=5.5cm}}
  \end{center}
\end{minipage}
\hfill
\begin{minipage}[t]{0.5\linewidth}
 \begin{center}
\begin{eqnarray}
\nonumber\;\; Q^2 &=& -q^2 = -(k' - k)^2 \\
\nonumber\;\; y &=& \frac{P \cdot q}{P \cdot k} \\
\nonumber\;\; x & = & \frac{Q^2}{2 P \cdot q}  \mathrm{ \;\;( Bjorken)} \\
\nonumber\;\; \nu & = & \frac{P\cdot q}{M_p} \\
\nonumber\;\; W^2 &=& (P')^2 = (P+q)^2 (=M_X^2)\\
\nonumber\;\;  s &=& (P+k)^2 \mathrm{( fixed)} 
\end{eqnarray}
 \end{center}
\end{minipage}
\label{fig:disdiag}
\caption{Lepton--hadron scattering: an exchange of a boson in the $t$--channel.}
\end{figure}

The lepton--hadron scattering is described in the framework  shown in figure~\ref{fig:disdiag}. The scattering can occur via the exchange of $\gamma$ or $Z$ bosons (neutral currents NC) or via $W$ bosons (charged currents CC). In the later case, a neutrino is expected in the final state. 
The incoming electron (with a four-momentum $k$) scatters off the proton ($P$) to a final state electron with four--momentum~$k'$ via a virtual photon $\gamma^*$ or a weak boson with a virtuality $Q^2$ in the $t$-channel. The Bjorken variable $x$ is associated with the fraction of the momentum of the proton carried by the struck parton. The total centre-of-mass energy is given by $\sqrt{s}$ and the energy of the  $\gamma^*p$ system is given by $W$, which is equivalent to the total mass of the hadronic system in the final state $M_X$. In the case of elastic scattering $M_X=M_p$ and from $M_X$ expression follows that $Q^2=2Pq$ and $x=1$ (the whole proton interacts).  Only two variables are independent, since the reaction is completely defined by the scatering angle and by the electron-parton centre-of-mass energy. 
The variable $\nu$ has a simple meaning in the proton rest frame, as the energy lost by the electron during the scattering $\nu=E_e-E_e'$, while $y$ represents the fractional energy loss $y=\frac{E_e-E_e'}{E_e} $. 
$Q^2$ can be expressed as a function of the electron energy and scattered angle $Q^2=4E_eE_e'\sin^2\frac{\theta}{2}$. 
From these relations, it is obvious that the DIS kinematics can be calculated from the measurement of the scattered electron only. The measurement of the hadrons in the final state, if available, can be exploited as an extra constraint or used in case of CC reactions, where the outgoing neutrino is not measured.

\section{The HERA project}
The idea for a large electron--proton collider to mark a new step in the studies for proton structure, beyond the fixed target experiments, was promoted already in seventies~\cite{Febel:1973ej}. The HERA collider project started in 1985 and produced  the first electron--proton collisions in 1992. It is composed of two accelerators designed to store and collide counter rotating electrons ($e^-$) or positrons ($e^+$) with an energy of 27.5~GeV and  protons with an energy of 920~GeV.  The operations came to an end in june 2007 and the final analyses using the collected data are in progress at present.
\par
HERA ring hosted two collider mode detectors H1 and ZEUS.
 They were build as hermetic ($4\pi$) multi-purpose detectors equipped with internal trackers able to measure charged particle momenta and calorimeters completing the measurement of the energy flow.   Two other experiments used $e^\pm$ or $p$ beams for fixed target studies: HERMES, dedicated to the study of polarised $e^\pm p(N)$ collisions and (until 2003) HERA-B dedicated to the study of beauty production in hadronic collisions. Since 2003, the $e^\pm$ beam were longitudinally polarised in collision mode with an average polarisation of  $P_{e^\pm}=30-40\%$.  An integrated luminosity of 300$^{-1}$ (200~pb$^{-1}$) has been  collected in $e^+p$ ($e^-p$) by each the two collider mode experiments, H1 and ZEUS. 
\par
\section{Deep-Inelastic Scattering Measurements at HERA}
 \begin{figure}[ht]
 \centerline{
 \epsfxsize=6.0cm\epsfysize=4.0cm\epsfbox{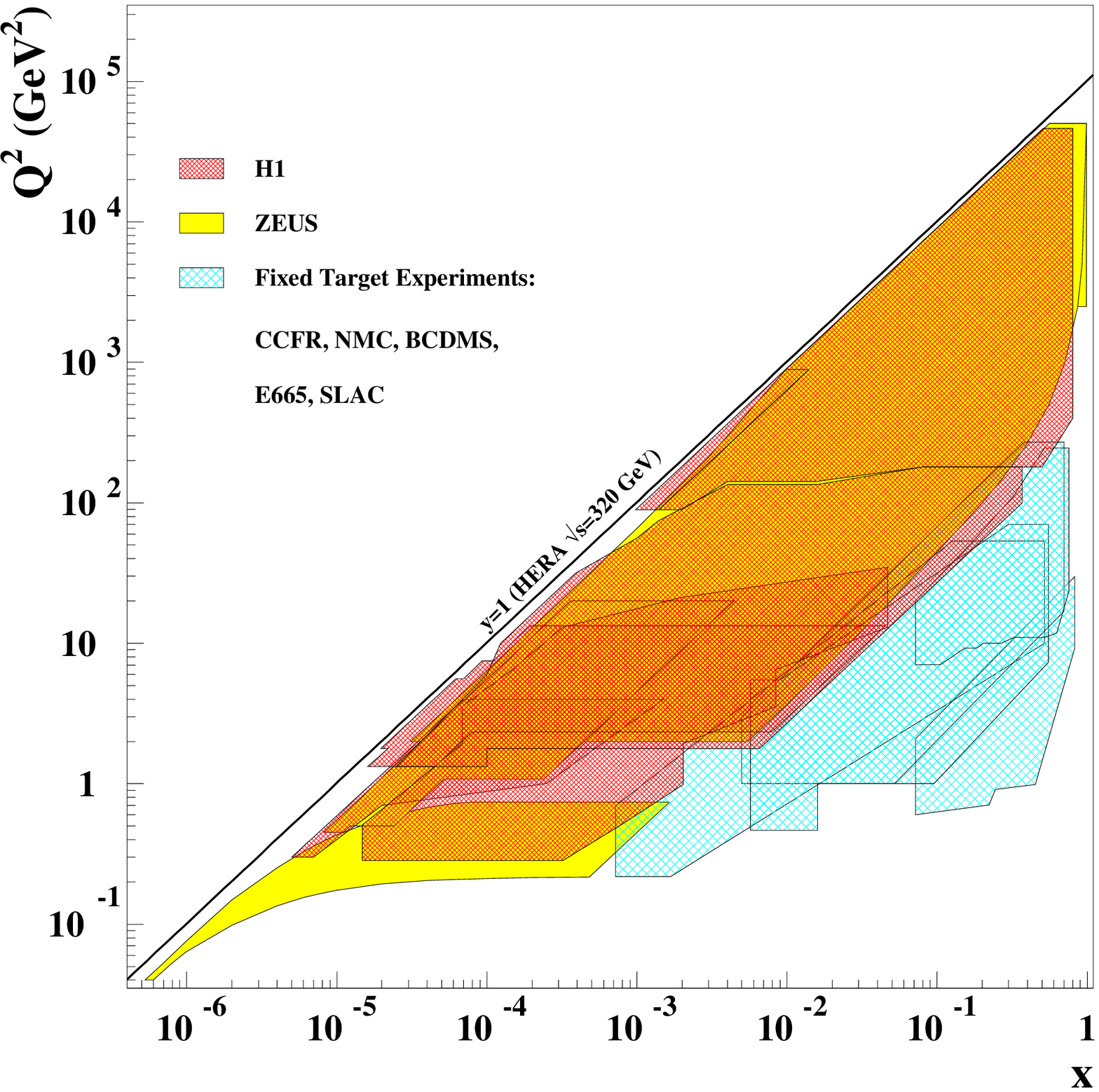}
 \epsfxsize=5.2cm\epsfbox{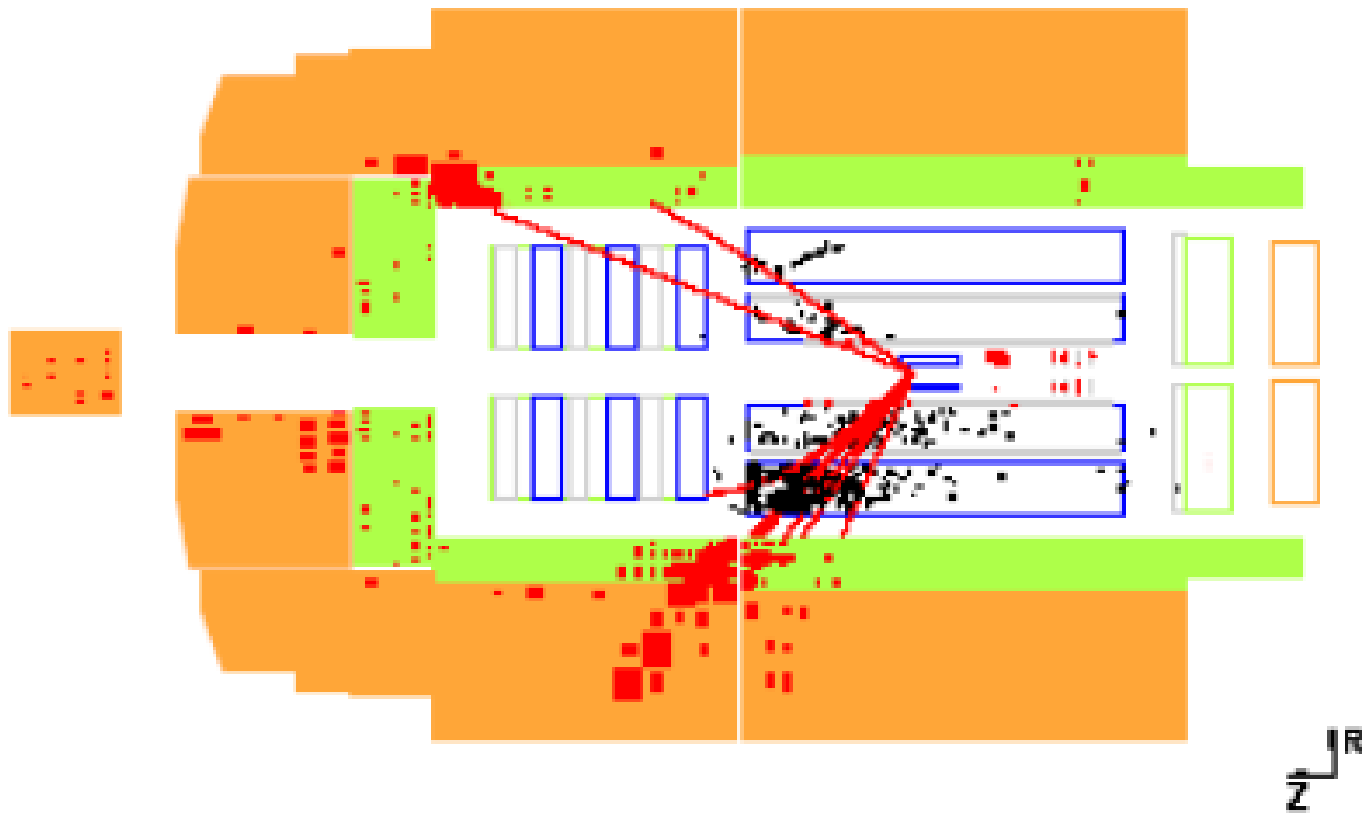}
 \epsfxsize=4.1cm\epsfbox{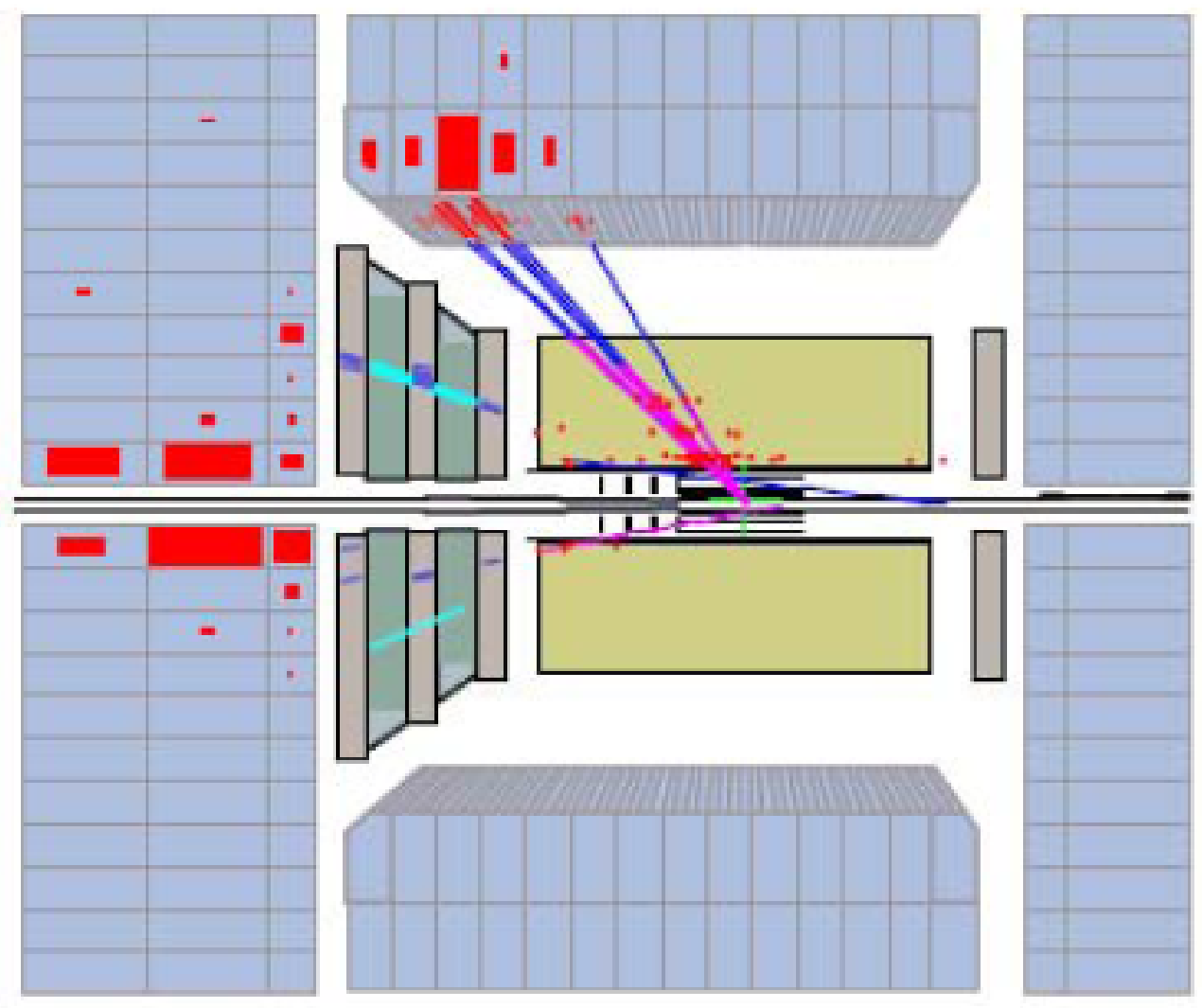}
 }
 \label{fig:ccncevents}
 \caption{ The  kinematic plane accessible at HERA compared to former fixed target experiments (left) and event displays of a neutral current scattering  event measured by H1 (center) and charged current scattering event measured by ZEUS (right).}
 \end{figure}
The H1 and ZEUS experiments measured  both neutral current (NC) and charged current (CC) processes. The kinematic $(x,Q^2)$ plane accessible at HERA is shown in figure~\ref{fig:ccncevents}~(left) with a $Q^2$ domain up to 50000~GeV$^{2}$ and $x$ down to $10^{-5}$. The NC events contain a prominent electron  and a jet of particles measured in the calorimeter, while in CC events only the jet is visible since the outgoing neutrino in not detected. Examples of such events are shown in figure~\ref{fig:ccncevents}.
\par
Since a large domain in $x$ and $Q^2$ is accessed, the NC cross section is sensitive to weak force  effects. The proton structure, as revealed by the photon and $Z^0$ boson in DIS, can be incorporated into the so-called generalised structure functions. The cross section is parameterised as following: 
\begin{equation}
\frac{\rm{d}^2\sigma^{\pm}_{NC}}{{\rm d}x{\rm d}Q^2}=
\frac{2\pi\alpha^2}{xQ^4}(Y_+\tilde{F}_2{\mp}Y_-x\tilde{F}_3-y^2\tilde{F}_L) 
\,\,\,,
\label{eq:ncxsec}
\end{equation}
The generalised structure functions $\Ftwo$ and $x\Fz$ can be further
decomposed as~\cite{klein}
\begin{eqnarray}
 \label{f2p}
 \Ftwo  \equiv & \Fem & - \ v_e  \ \frac{\kappa  \QQ}{(\QQ + M_Z^2)}
  \Fint  \,\,\, + (v_e^2+a_e^2)  
 \left(\frac{\kappa  Q^2}{\QQ + M_Z^2}\right)^2 \Fwk\,, \\
 \label{f3p}
 x\Fz    \equiv &      & - \ a_e  \ \frac{\kappa  \QQ}{(\QQ + M_Z^2)} 
 x\Fzint + \,\, (2 v_e a_e) \,\,
 \left(\frac{\kappa  Q^2}{\QQ + M_Z^2}\right)^2  x\Fzwk\,,
\end{eqnarray} 
with $\kappa^{-1}=4\frac{M_W^2}{M_Z^2}(1-\frac{M_W^2}{M_Z^2})$ in the
on-mass-shell scheme~\cite{pdg}.  The quantities $v_e$ and $a_e$ are the
vector and axial-vector weak couplings of the electron or positron
 to the
$Z^{0}$~\cite{pdg}.  The electromagnetic structure function $\Fem$
originates from photon exchange only
 and dominates over the vast majority of the measured phase space.
 The functions $\Fwk$ and $x \Fzwk$
 are the contributions to $\Ftwo$ and $x\Fz$ from $Z^0$ exchange and the
 functions $\Fint$ and $x\Fzint$ are the contributions from $\gamma Z$
 interference. These contributions are significant only at high $Q^2$.  The structure functions provide however direct information on the proton components. Within the so-called quark-parton model (QPM), the proton is composed of spin one half partons, called quarks $q=u,d$ with fractionary charges $e_q$. This model predicts the structure functions as combinations of quark densities $q(x)$. For instance $F_2=\sum_{q} e_q q(x)$ and is independent of $Q^2$ (Bjorken scaling).  This model is improved within the Quantum Chromodynamics (QCD) predicting quark interactions via gluons, the careers of the strong force, and leading to $F_2$ dependence on $Q^2$ (scaling violation).
\begin{figure}
\centerline{
\epsfxsize=6.0cm\epsfysize=7.0cm\epsfbox{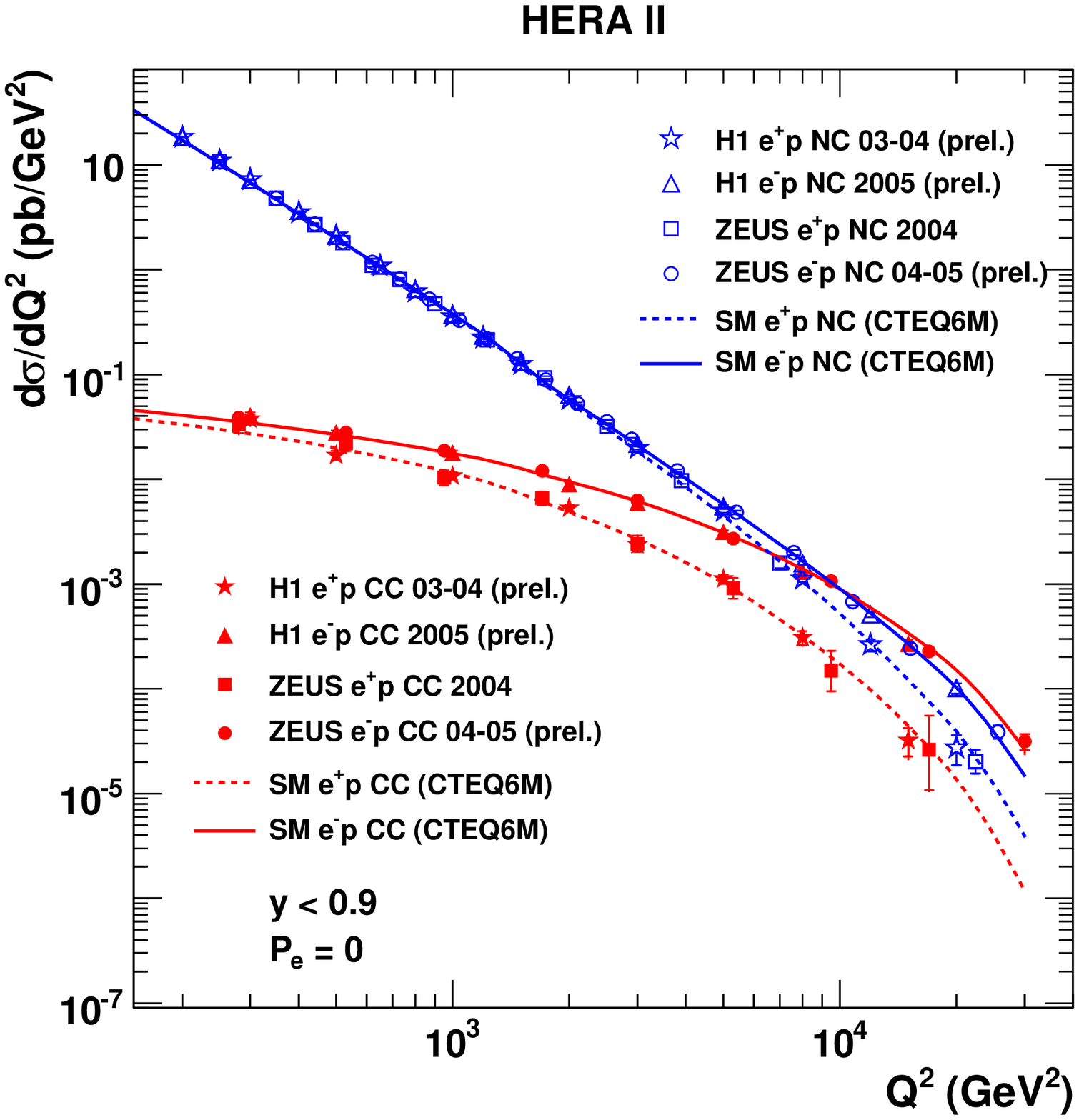}
\epsfxsize=7.0cm\epsfbox{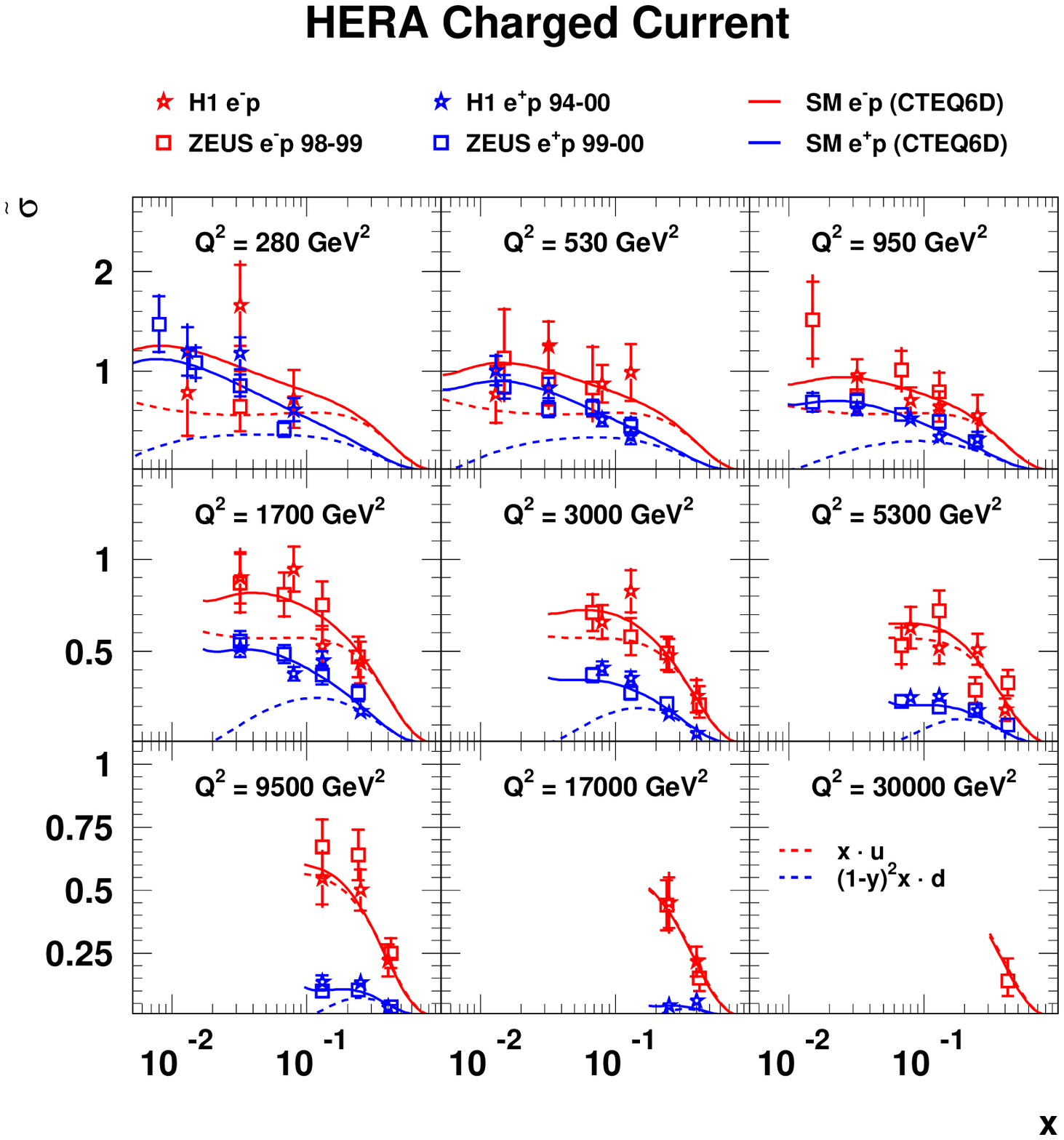} }
\caption{Left: the charged current and neutral current cross section as a function of $Q^2$ measured in $e^\pm p$ collisions at HERA. Right: the charged current reduced cross section $\tilde{\sigma}_{CC}$ as a function of $x$ for various $Q^2$ values measured in electron-- and positron--proton collisions. }
\label{fig:ccnc}
\end{figure}
 \par
The charged current (CC) interactions, 
$e^\pm p \rightarrow \overline{\nu}_e^{\mbox{\tiny
\hspace{-3mm}\raisebox{0.3mm}{(}\hspace{2.5mm}\raisebox{0.3mm}{)}}}X$,
are mediated by the exchange of a $W$ boson in the $t$ channel.  The cross section is parameterised as:
\begin{equation}
\nonumber \frac{d^2\sigma^{\rm CC}(e^\pm p)}{dxdQ^2} = 
\frac{G^2_F}{2\pi x}\left[\frac{M^2_W}{M^2_W+Q^2}\right]^2\tilde{\sigma}^\pm_{CC}(x,Q^2)
\,,\label{eqn:xscc}
\end{equation}
with $\tilde{\sigma}^\pm_{CC}(x,Q^2)=
\frac{1}{2}\left[Y_+W_2^\pm (x,Q^2)\mp Y_-xW_3^\pm (x,Q^2)-y^2W_L^\pm (x,Q^2)\right]\,$.
$\tilde{\sigma}$ is the reduced cross section, $G_F$ is the  Fermi constant, $M_W$, 
the mass of the $W$ boson, and $W_2$,
$xW_3$ and $W_L$, CC structure functions.
In the quark parton model, 
the structure functions $W^\pm_2$ and $xW^\pm_3$ may be interpreted as
lepton charge dependent sums and differences of quark and anti-quark 
distributions:
$W^+_2=x(\overline{U}+D),\hspace{3mm} xW^+_3=x(D-\overline{U}),\hspace{3mm}
W^-_2=x(U+\overline{D}),\hspace{3mm} xW^-_3=x(U-\overline{D})\,$,
whereas $W^\pm_L=0$. 
The terms $xU$, $xD$, $x\overline{U}$ and $x\overline{D}$
are defined as the sum of up-type, of down-type and of their anti-quark-type 
distributions.
\par
The differential NC and CC cross sections as a function of $Q^2$ are shown in figure~\ref{fig:ccnc}~(left) for $e^\pm p$ collisions. At low $Q^2$ the NC cross section, driven by the electromagnetic interaction, is two orders of magnitude larger than the CC cross section which correspond to a pure weak interaction. At large $Q^2\sim M_{W,Z}^2$ the two cross sections are similar, which can be interpreted as a hint for electroweak unification. The largest $Q^2$ measurement corresponds to a resolution $\delta~\simeq 10^{-18}$~m, i.e. 1/1000 the proton size. The agreement between the measurement and the prediction based on improved parton model constitutes a spectacular confirmation of QCD and suggests that no evidence for quark substructure is observed at present.
\par
The double differential reduced cross section $\tilde{\sigma}_{CC}(x,Q^2)$ is shown in figure~\ref{fig:ccnc}~(right). 
The CC processes are sensitive to individual quark flavours, especially visible at large $Q^2$: the $e^+p$ collisions probe the $d$ quark distribution, while $e^-p$ are more sensitive to the $u$ distribution. This is a very useful feature of the CC processes compared to the NC, where the quark flavour separation is weaker.

\section{Structure functions measurements: $F_2$, $F_L$ and $xF_3$}
\begin{figure}
\centerline{
\epsfxsize=12.2cm\epsfysize=7.5cm\epsfbox{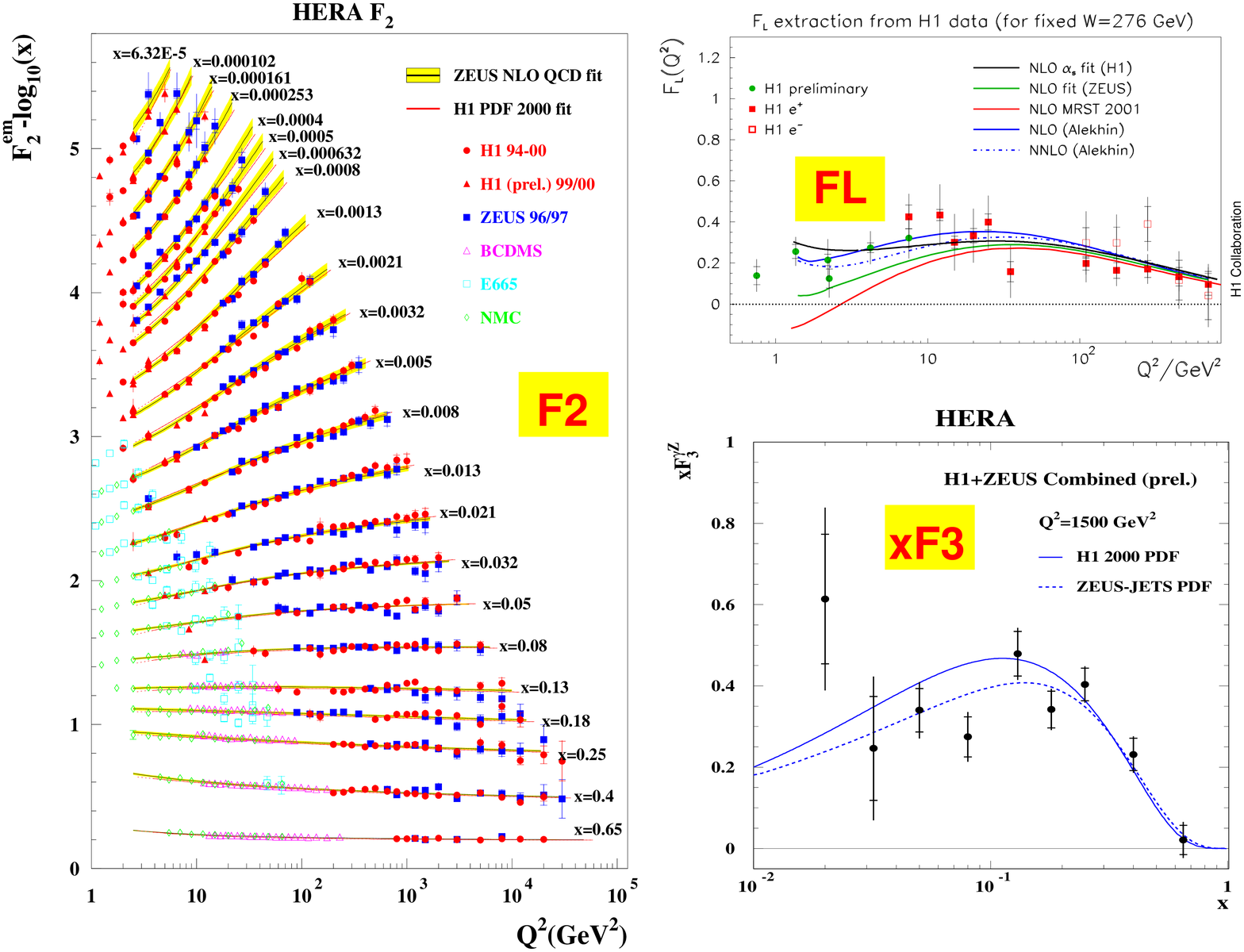}
}
\caption{The determination of the structure functions $F_2$, $F_L$ and $xF_3$ from HERA measurements.}
\label{fig:SF}
\end{figure}
The NC cross section is dominated over a large domain by the $F_2$ contributions, defined in equation~\ref{eq:ncxsec}. The measurement of the NC cross section at HERA can therefore be translated into an $F_2$ measurement, which is shown in figure~\ref{fig:SF} together with the previous measurements performed at
 fixed target experiments. One can observe the Bjorken scaling in the region at high $x$, but obvious scaling violation at lower $x$. This can be understood in terms of DGLAP equations~\cite{DGLAP} as a contribution driven by the gluon $\partial F_2(x,Q^2)/\partial \ln (Q^2) \approx (10\,\alpha_s(Q^2) / 27\pi)\,xg(x,Q^2)\,$.
while at high $x$ the quark pdf's are the major contributors to the evolution in DGLAP equations.
\par
\begin{figure}
\centerline{
\epsfxsize=6.5cm\epsfbox{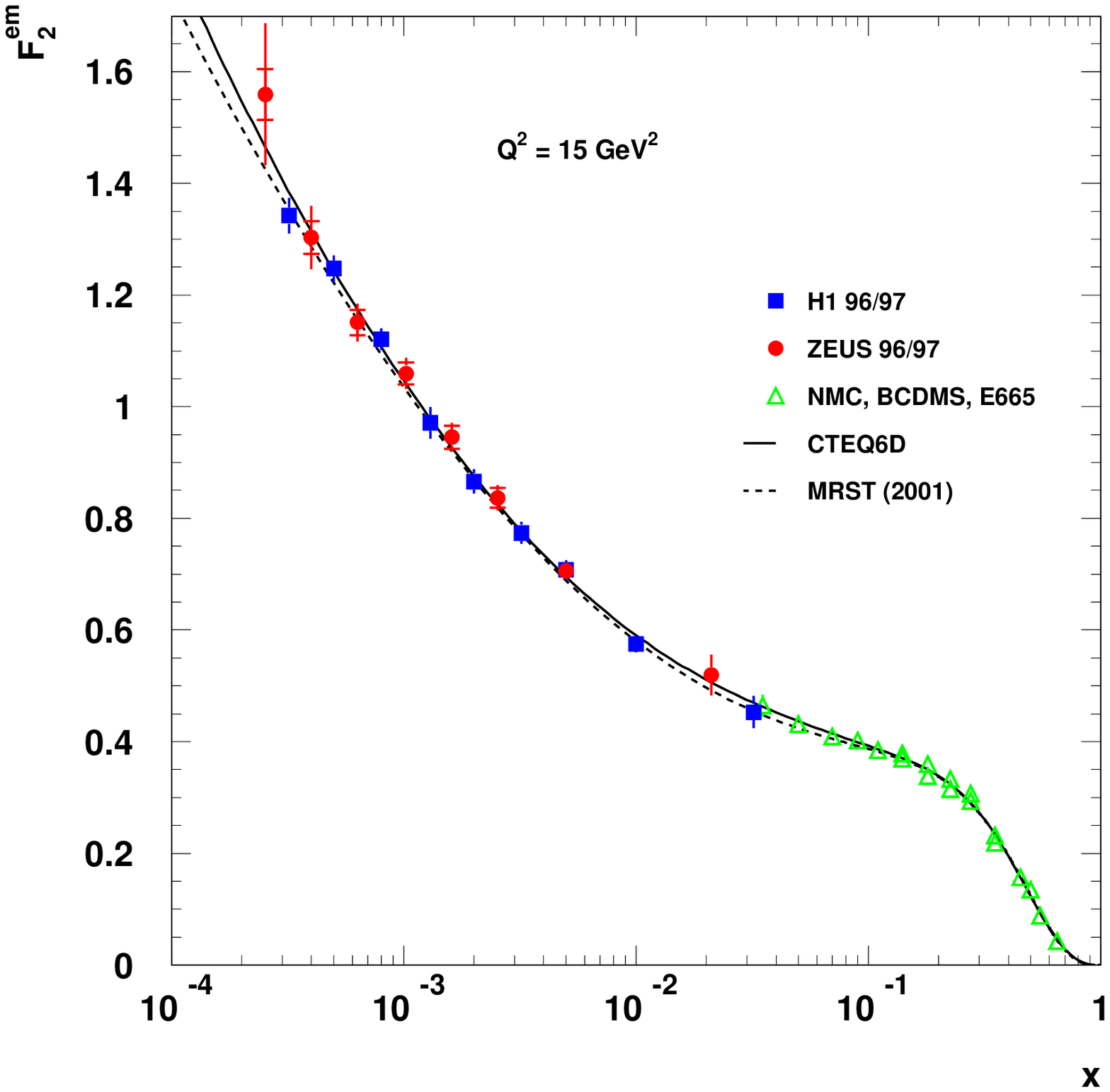}
\epsfxsize=5.3cm\epsfysize=6.0cm\epsfbox{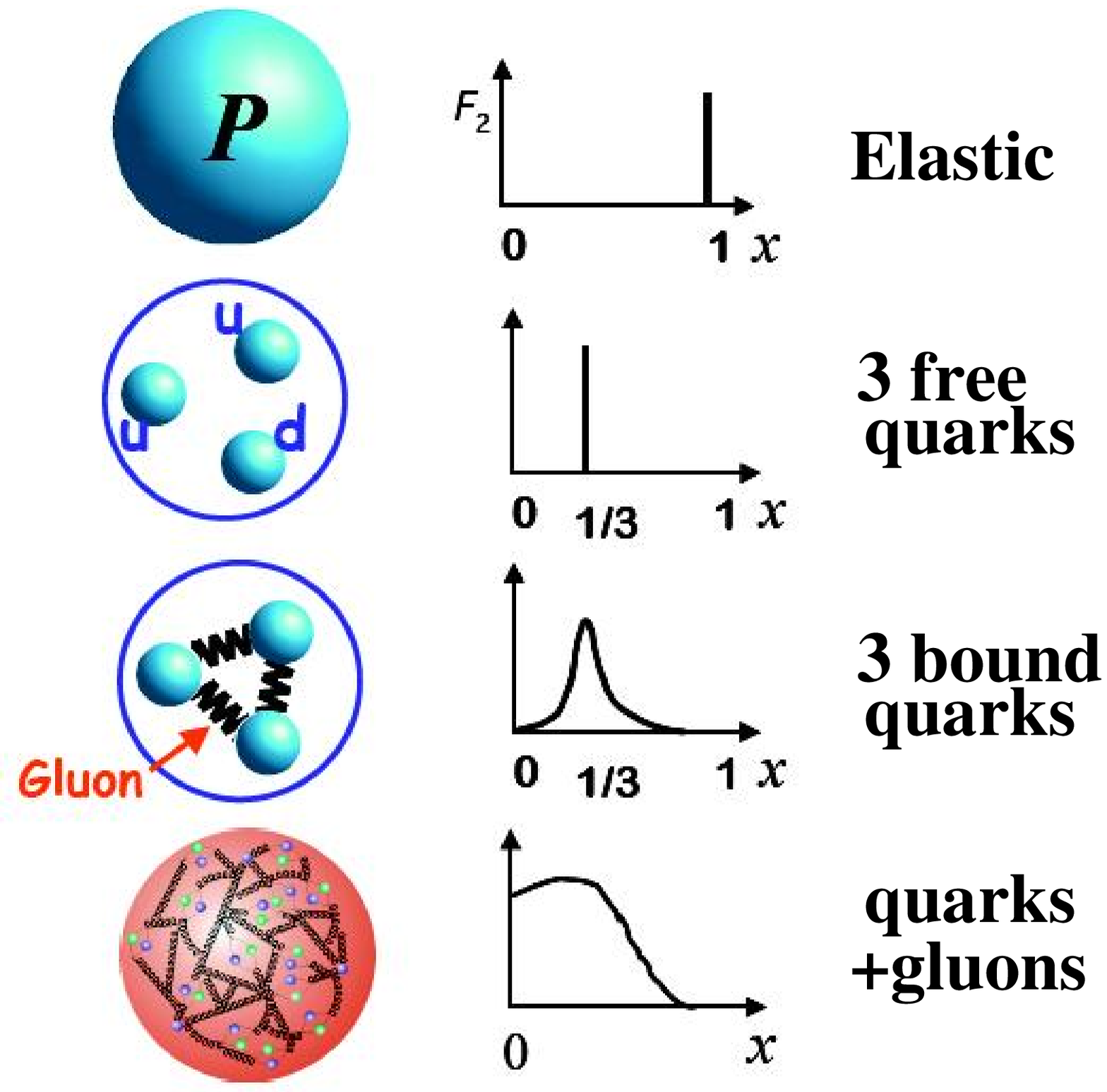}
}
\caption{Left: The measurement of $F_2$ as a function of $x$ for $Q^2=15$~GeV. Right: a sketch of the correspondence between $F_2$ shape as a function of $x$ and the quark-parton model.}
\label{fig:lowx}
\end{figure}
From the $F_2$ measurements at fixed $Q^2$ one can observe a steep increase of $F_2$ towards low $x$, as shown in figure~\ref{fig:lowx}~(left). The region at low $x$ is populated by quarks which have undergone a hard or multiple gluon radiation and carry a low fraction of the proton momentum at the time of the interaction. The observation of such large fluctuations to very high parton density is driven by the uncertainty principle, which requires that the interaction time be very short and therefore at high $Q^2$. In this regime, it is expected that the structure function grows at low $x$ and shrinks at large $x$, confirmed by the experimental observation.
The rise of the structure functions at low $x$ is one of the most surprising observations at HERA. It is predicted in the double leading log limit of QCD~\cite{DLL}. It can be intuitively understood in terms of gluon driven parton production at low $x$, as depicted in figure~\ref{fig:lowx}~(right). 
\par
The longitudinal structure function  $F_L$ is usually a small correction, only visible at large $y$. The $F_L$ measurement from the cross section has to proceed in such a way that $F_2$ contribution is separated. Indirect methods assume some parameterisation of $F_2$ to extract $F_L$. 
Using this method, an $F_L$ determination can be performed and is shown in figure~\ref{fig:SF} at fixed $W$ (the $\gamma^* p$ centre-of-mass energy). 
In the naive QPM the longitudinal structure function $F_L =F_2-2xF_1 \equiv 0$ and therefore $F_L$ contains by definition the deviations from the Callan-Gross relation. It can be shown that $F_L$ is directly related to the gluon density in the proton~\cite{Altarelli:1978tq,Cooper-Sarkar:1987ds} $xg(x) = 1.8[ \frac{3 \pi}{2 \alpha_s} F_L(0.4 x) - F_2(0.8 x] \simeq \frac{8.3}{\alpha_s} F_L$  meaning that at low $x$, to a good approximation $F_L$ is a direct measure for the  gluon distribution.
\par
A direct measurement of $F_L$ can be performed if the cross section $\sigma  \sim F_2 (x,Q^2) + f(y) \; F_L  (x,Q^2) $ is measured at fixed $x$ and $Q^2$ but variable $y$. This can only be performed if the centre-of-mass energy is varied, for instance by reducing the proton beam energy. Eliminating $F2(x,Q2)$, $F_L$ can be directly measured with reduced uncertainties from the difference of cross sections: $\displaystyle F_L \sim C(y)*(\sigma(E_p^1) -\sigma(E_p^2)) $.
 Measurements of DIS at HERA at lower proton energies of 460 GeV and 575 GeV has been performed at the end of the run in 2007 in order to perform the first direct measurement of $F_L$ in the low $x$ and high $Q^2$ regime.
\par
The structure function $x{\tilde F}_3$ can be obtained from
the cross section difference between electron and positron unpolarised data
$x{\tilde F}_3 = \frac{Y_+}{2Y_-}\left[\tilde{\sigma}^-(x,Q^2)-
\tilde{\sigma}^+(x,Q^2)\right]$
The dominant contribution to $xF_3$ arises from the $\gamma Z$ interference.
 In leading order QCD the interference structure function $xF_3^{\gamma Z}$ 
 can be written as
 $ xF_3^{\gamma Z} = 2x [e_u a_u (U-\overline{U}) + e_d a_d (D - \overline{D})]$,
 with $U=u+c$ and $D=d+s$ thus provides information  about the light quark
 axial vector couplings ($a_u,\,a_d$) and the sign of the
 electric quark charges ($e_u,\,e_d$). The averaged $xF_3^{\gamma Z}$,
determined by H1 and ZEUS for a $Q^2$ value of $1500{\rm\,GeV}^2$, is shown
in figure~\ref{fig:SF}. 

\section{Parton distribution functions and electroweak effects}
\begin{figure}
\centerline{
\epsfxsize=6.5cm\epsfbox{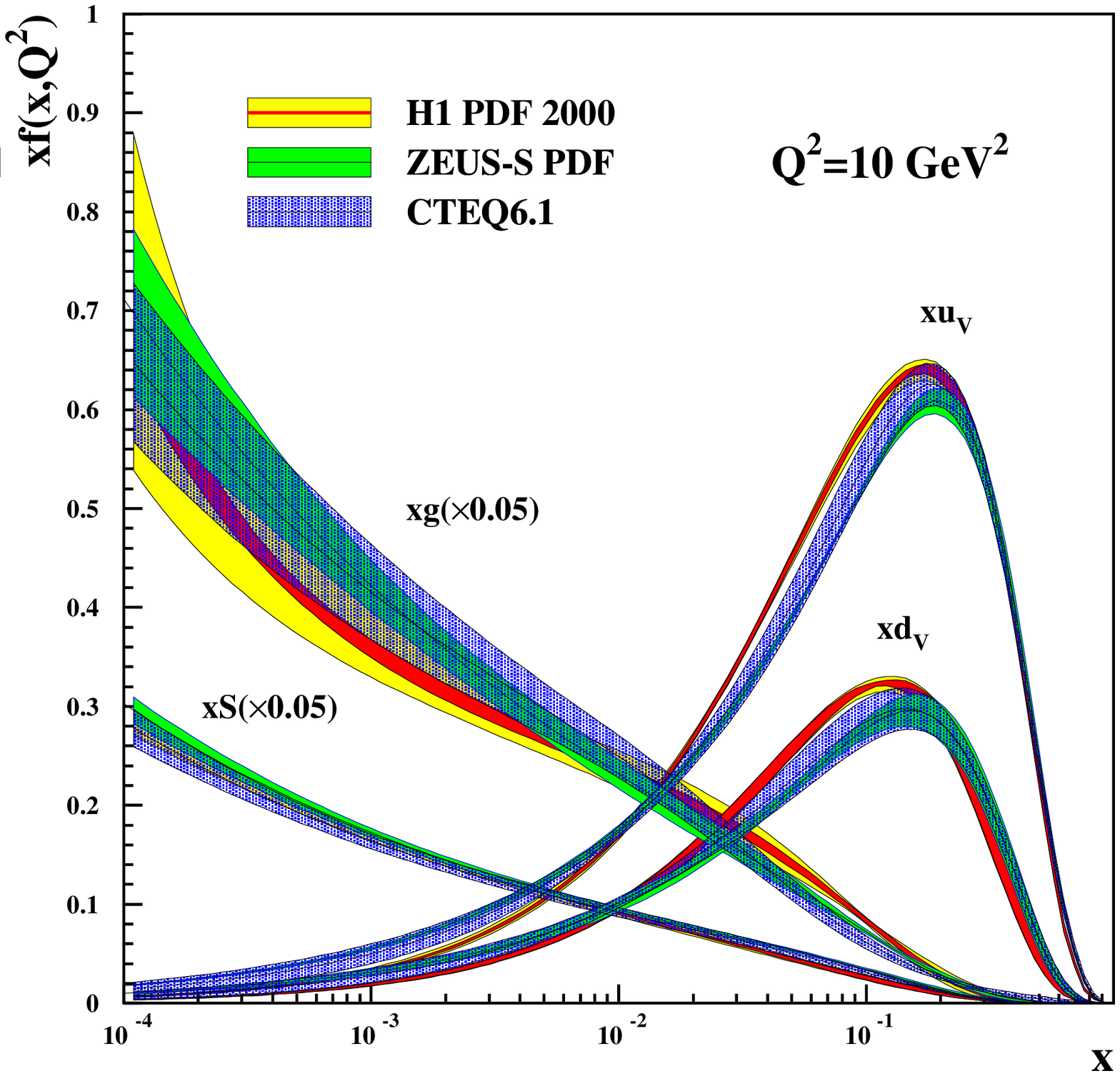}
\epsfxsize=6.5cm\epsfbox{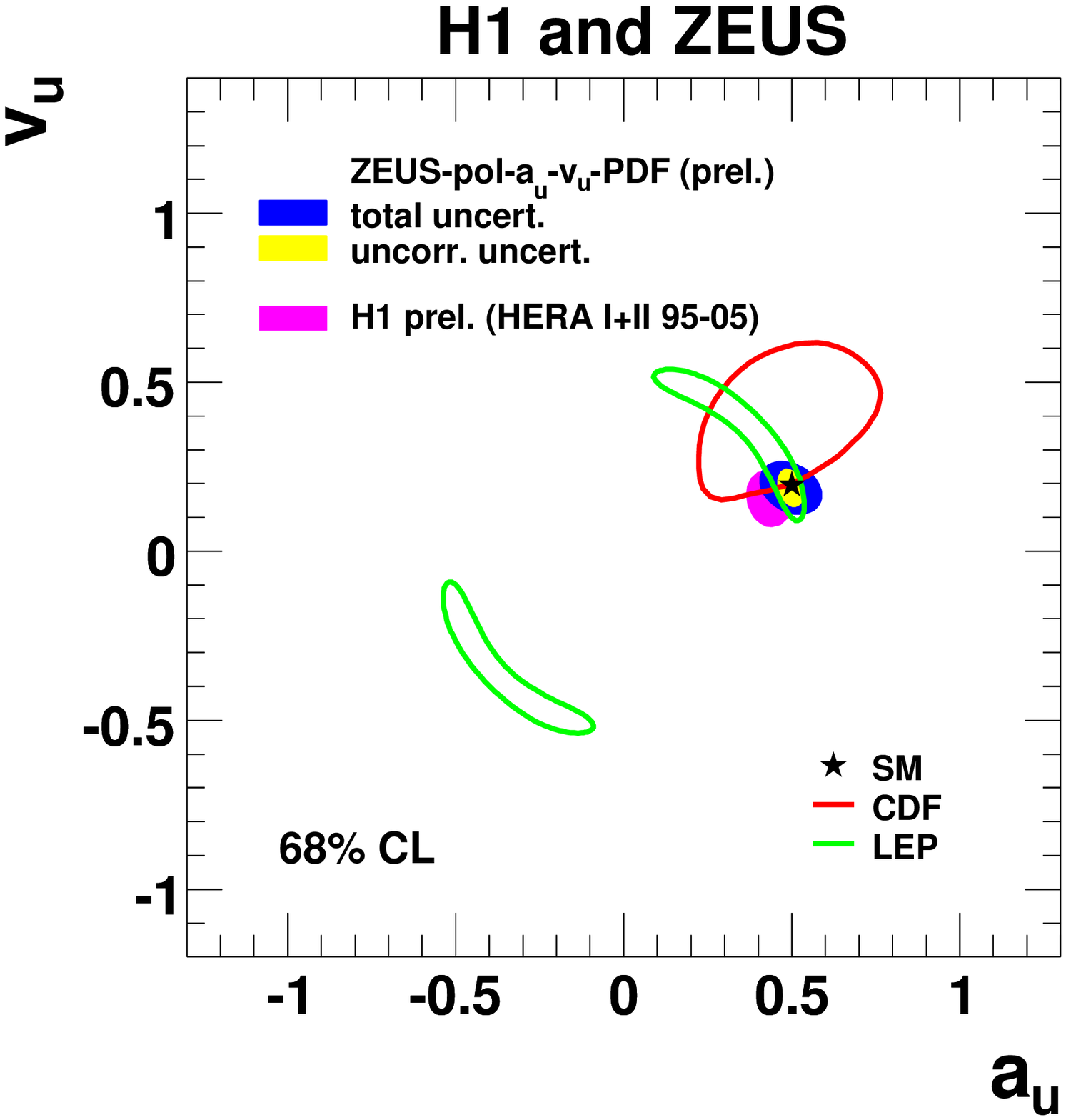}}
\caption{Left: The parton distribution functions extracted from HERA data. Right: Axial and vector couplings of the $u$--quark measured from the combined electroweak--QCD fit at HERA and compared with measurements from LEP (using light quarks production at Z p\^ole $e^+e^- \rightarrow q\bar{q}$) and Tevatron (from Drell-Yan electron pair production $q\bar{q} \rightarrow e^+e^- $).}
\label{fig:pdfs}
\end{figure}
The NC and CC cross section measurements are used in a global fit in order to extract the parton distribution functions (pdf's)~\cite{Adloff:2003uh,Chekanov:2002pv}. The shapes for the quarks $q(x,Q_0^2)$ and gluon $g(x,Q_0^2)$ distributions are parametrised as a function of $x$ at a given scale $Q_0^2$ and evolved using DGLAP equations~\cite{DGLAP} to each $(x,Q^2)$ point where the cross section has been measured. The theoretical cross section can therefore be accurately calculated as a function of the pdf's parameters. A $\chi^2$  is then built using the measurements and the predictions for all measurements points and minimised to extract the non-perturbative pdf's parameters. Since the number of parameters (typically 10) is much lower than the number of measurements (several hundred) the fit also consitutes  a very powerful test of QCD. The structure functions from the fit are compared with data in figure~\ref{fig:SF}. The parton distribution functions are extracted using the decomposition of the structure function described above. As an example, the pdf's obtained for $Q^2=10$~GeV$^2$ are shown in figure~\ref{fig:pdfs}. The valence distributions peak at $1/3$ as expected from simple counting with $u_V$ twice as large as $d_V$. Gluon distribution is enhanced at low $x$. The knowledge of the proton structure deduced from inclusive CC/NC measurements can be used to calculate the cross section of exclusive processes leading to a specific final state $FS$ as a convolution of the parton level cross section with pdf's, for instance: $\sigma_{ep\rightarrow FS} = \sigma_{\rm eq -> FS} \otimes q(x,Q^2)$. This factorisation can also be used to calculate the cross section of processes produced in proton--proton collisions using the pdf's measured in DIS.   
\par
Recently, a new approach has been adopted by the H1 and ZEUS collaborations\cite{h1zeus:ewfit}, performing a combined QCD--electroweak fit.  The strategy is to leave free in the fit the EW parameters together with the  parameterisation of the parton distribution functions.  
Due to the $t$-channel electron-quark scattering via $Z^0$ bosons, the DIS cross sections at high $Q^2$ are sensitive to light quark axial ($\mathrm{a}_q$) and vector ($\mathrm{v}_q$) coupling to the $Z^0$. This dependence includes linear terms with significant weight in the cross section which allow to determine not only the  value but also the sign of the couplings. 
The measurements of the $u$--quark couplings obtained at HERA, LEP and Tevatron are shown in figure~\ref{fig:pdfs}. 
\section{$e^\pm p$ collision with a polarised lepton beam}
\begin{figure}
\centerline{
\epsfxsize=6.0cm\epsfysize=5.5cm\epsfbox{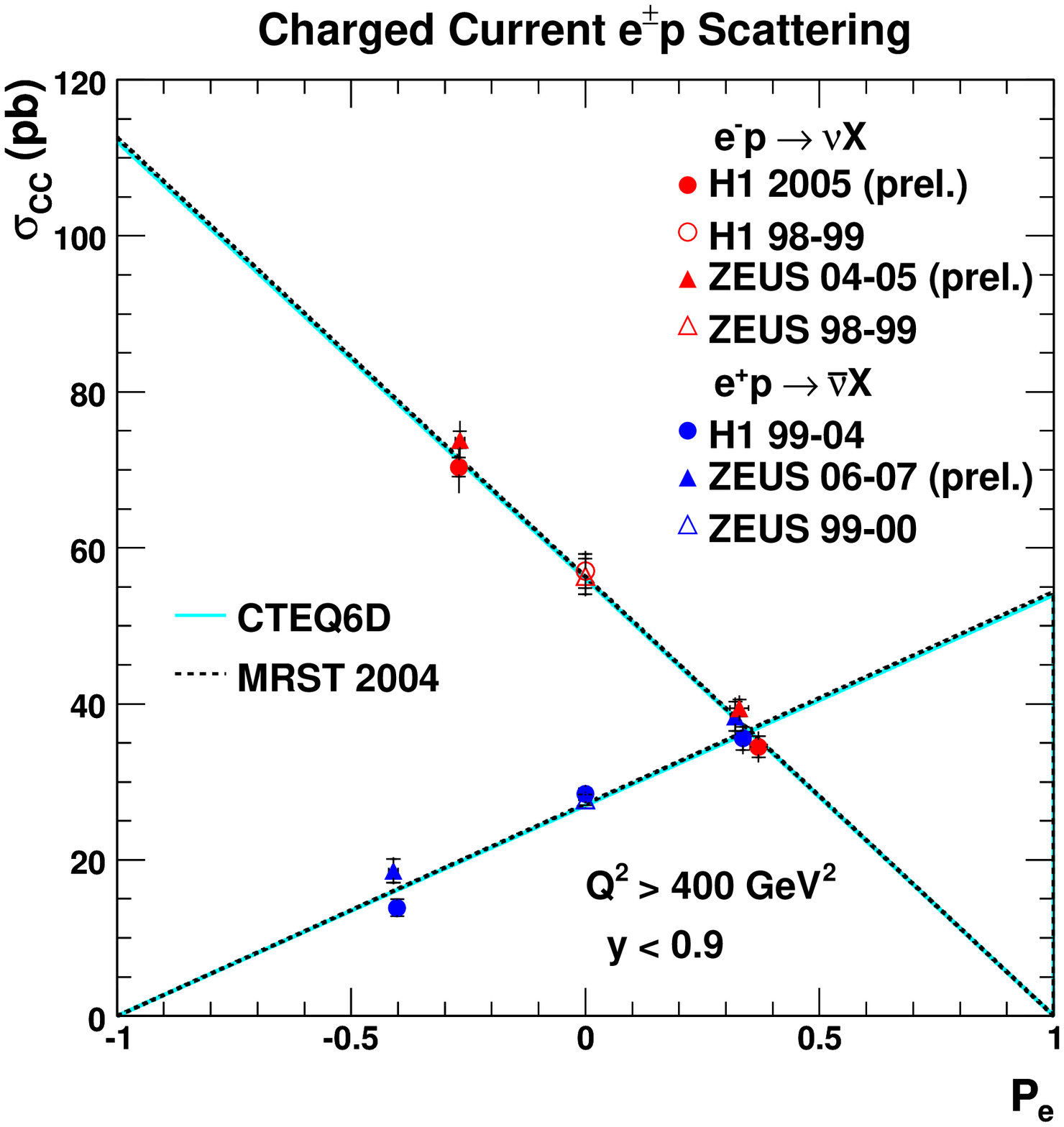}
\epsfxsize=6.0cm\epsfysize=5.5cm\epsfbox{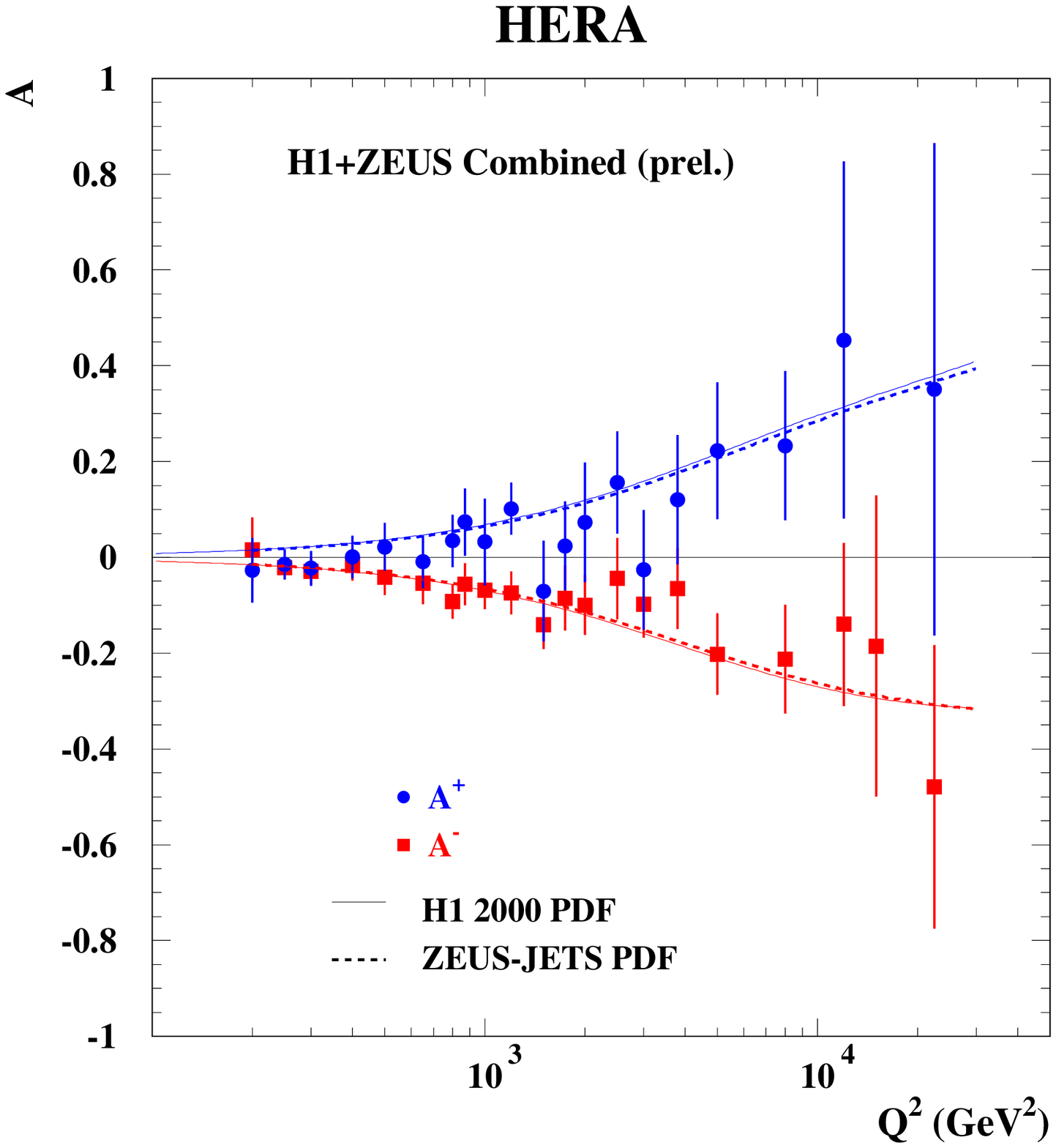}
}
\caption{Left: The dependence of the charged current cross section on  the electron or positron beam polarisation at HERA. Right: The polarisation asymmetry of the NC cross section at HERA.}
\label{fig:ccncpol}
\end{figure}

The polarisation of the electron beam at HERA II allows a test of the parity non-conservation effects typical of the electroweak sector. 
The most prominent effect is predicted in the CC process, for which the cross section depends linearly on the $e^\pm$--beam polarisation: $\mathrm{\sigma^{e^{{\pm}} p} (P) =(1 {\pm} P) \sigma^{e^{{\pm}} p}_{P=0}}$. The results\cite{ccpol} obtained for the first time in $e^{\pm}p$ collisions are shown in figure~\ref{fig:ccncpol}. The expected linear dependence is confirmed and provides supporting evidence for the V-A structure of charged currents in the Standard Model.
\par
Due to parity violating couplings of the $Z$ boson, the $e^\pm$ beam polarisation effects can also be measured in NC processes at high $Q^2$. 
The charge dependent longitudinal polarisation
asymmetries of the neutral current cross sections, defined as
\begin{equation} \label{apm}
 \nonumber A^\pm = \frac{2}{P_R-P_L} \cdot \frac{\sigma^{\pm}(P_R) -\sigma^{\pm}(P_L)}
                          {\sigma^{\pm}(P_R) +\sigma^{\pm}(P_L)} \simeq \mp  k a_e \frac{F_2^{\gamma Z}}{F_2},
\end{equation}
measure to a very good approximation the structure function ratio. 
These asymmetries are proportional to combinations $\mathrm{a}_e \mathrm{v}_q$ and 
thus provide a direct measure of parity violation.
In the Standard Model $A^+$ is expected
to be positive and about equal to $-A^-$. 
At large $x$  the asymmetries measure 
the $d/u$ ratio of the valence quark distributions according to
 $A^\pm   \simeq  \pm  k \frac{1+d_v/u_v}{4+d_v/u_v}.$
The measurement from ZEUS and H1~\cite{nc_zeush1}, shown in figure~\ref{fig:ccncpol}, are in agreement with the theoretical predictions.
\section{Outlook}
The study of deep-inelastic scattering is a fundamental branch of high energy physics. The structure of matter was last time resolved to new components, the quarks, in the first break-up of the proton at SLAC in 1968. Since then, the Standard Model of particle physics has become a well established theory with the last quark, the top,  discovered in 1994. The knowledge of the structure of the baryonic matter, dominating the visible universe, has made  huge progress in the last decades, thanks to an impressive effort to unravel the nucleon structure in fixed target experiments and at the HERA $ep$ collider.  
The knowledge acquired at HERA is invaluable also for the physics of the Large Hadron Collider, foreseen to start $pp$ collisions at 14~TeV in 2008. 
 By enabling even more ambitious DIS experiments~\cite{Dainton:2006wd} beyond HERA, the matter structure investigations may gain a new momentum.

\doingARLO[\bibliographystyle{aipproc}]
          {\ifthenelse{\equal{\AIPcitestyleselect}{num}}
             {\bibliographystyle{arlonum}}
             {\bibliographystyle{arlobib}}
          }

\hyphenation{Post-Script Sprin-ger}

\end{document}